\documentclass[12pt]{article}
\pdfoutput=1
\usepackage{tikz}
\usetikzlibrary{matrix,arrows,decorations.pathmorphing,decorations.pathreplacing}
\usepackage{jheppub}
\usepackage{amsmath,amssymb,euscript,array,mathrsfs,appendix,ctable}
\usepackage{arydshln}
\usepackage{todonotes}
\usepackage{graphicx}
\usepackage[normalem]{ulem}

\newcommand\blank[1]{#1}
\renewcommand\blank[1]{}

\def\mpsu{\mathfrak{psu}}

\def\msl{\mathfrak{sl}}

\def\msl{\mathfrak {sl}}

\def\mh{\mathfrak h}

\def\msu{\mathfrak{su}}
\def\msl{\mathfrak{sl}}

\def\n{\nu}

\newcommand{\IM}{\operatorname{Im}}

\def\B0{{\boldsymbol 0}}

\def\R{{\mathbb R}}

\def\Dbarslash{\,\,{\raise.15ex\hbox{/}\mkern-12mu {\bar D}}}
\def\Dslash{\,\,{\raise.15ex\hbox{/}\mkern-12mu D}}
\def\delslash{\,\,{\raise.15ex\hbox{/}\mkern-9mu \partial}}
\def\delbarslash{\,\,{\raise.15ex\hbox{/}\mkern-9mu {\bar\partial}}}

\def\ket#1{\left| #1\right\rangle}

\newcommand{\EQ}[1]{\begin{equation}\begin{split} #1
\end{split}\end{equation}}

\title{Bound States of the ${\boldsymbol q}$-Deformed \text{AdS}$_{\boldsymbol 5}{\boldsymbol\times}$S$^{\boldsymbol 5}$ Superstring S-matrix}
\author[a]{Ben Hoare,}
\author[b]{Timothy J. Hollowood}
\author[c]{and J. Luis Miramontes}
\affiliation[a]{Theoretical Physics Group, Blackett Laboratory, Imperial College, London SW7 2AZ, U.K.}
\affiliation[b]{Department of Physics, Swansea University, Swansea, SA2 8PP, U.K.}
\affiliation[c]{Departamento de F\'\i sica de Part\'\i culas and IGFAE,
Universidad de Santiago de Compostela, 15782 Santiago de Compostela, Spain}

\emailAdd{benjamin.hoare08@imperial.ac.uk} 
\emailAdd{t.hollowood@swansea.ac.uk}
\emailAdd{jluis.miramontes@usc.es}

\abstract{The investigation of the $q$ deformation of the S-matrix for excitations on the string world sheet in $\text{\text{AdS}}_5\times S^5$ is continued. We argue that due to the lack of Lorentz invariance the situation is more subtle than in a relativistic theory in that the nature of bound states depends on their momentum. At low enough momentum $|p|<E$ the bound states transform in the anti-symmetric representation of the super-algebra symmetry and become the solitons of the Pohlmeyer reduced theory in the relativistic limit. At a critical momentum $|p|=E$ they become marginally unstable, and at higher momenta the stable bound states are in the symmetric representation and become the familiar magnons in the string limit as $q\to1$. This subtlety fixes a problem involving the consistency of crossing symmetry with the relativistic limit found in earlier work. With mirror kinematics, obtained after a double Wick rotation, the bound state structure is simpler and there are no marginally 
unstable bound states. }

\notoc
\begin{document}

\begin{flushright}   \small Imperial/TP/12/BH/01
\end{flushright}
\vspace{0.5cm} 

\maketitle

\newpage

\section{Introduction}

This note is a companion to \cite{Hoare:2011wr}. In \cite{Hoare:2011wr} we considered a certain deformation of the S-matrix for excitations on the superstring worldsheet propagating in $\text{\text{AdS}}_5\times S^5$. This deformed S-matrix theory is based on the fundamental R-matrix of the quantum-deformed Hubbard chain~\cite{Beisert:2008tw,arXiv:1002.1097}. The scattering theory depends on two separate couplings: $g$, the original coupling of the magnon theory that corresponds to the 't~Hooft coupling of the dual gauge theory, and a new quantum deformation parameter $q=e^{i \pi/k}$. 
 
In \cite{Hoare:2011wr} we showed that there are two versions of the scattering theory, one appropriate for the original magnon theory and the other for the so-called mirror theory~\cite{arXiv:0710.1568}. The latter is related to the former by a double Wick rotation and is the appropriate theory for setting up the thermodynamic Bethe Ansatz.\footnote{In fact in \cite{Hoare:2011wr} we constructed four different S-matrices. However, in the present work we are only interested in the two built from the dressing factor $\sigma$ and not the other two built from the alternative dressing factor $\widehat\sigma$.} In a relativistic theory a double Wick rotation leads to the same theory, but for the non-relativistic situation that ensues with general $(g,k)$ these are distinct theories. 

The bound state R-matrix of the deformed theory was constructed in \cite{deLeeuw:2011jr} using the underlying infinite dimensional quantum affine symmetry \cite{Beisert:2011wq,deLeeuw:2011fr,deLeeuw:2012jf}. Together with the dressing phase for the fundamental S-matrix constructed in \cite{Hoare:2011wr} one should be able to construct the exact S-matrix. However, due to the non-relativistic nature of the deformed theory, the identification of the bound state spectrum is subtle and depends on the couplings and the momenta of the particles.

The spectrum of the non-relativistic $q$-deformed, or interpolating, theory has some unusual properties that we conceptualize below:
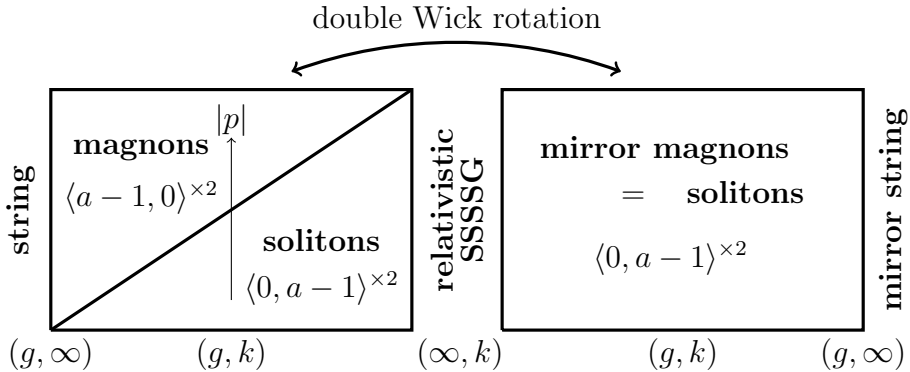
\begin{center}
\begin{tikzpicture}[scale=0.8]
\draw[-,very thick] (0,0) -- (6,0) -- (6,4) -- (0,4) -- (0,0);
\draw[-,very thick] (0,0) -- (6,4);
\node at (0,-0.4) (a1) {$(g,\infty)$};
\node at (6.8,-0.4) (a2) {$(\infty,k)$};
\node at (3,-0.4) (a3) {$(g,k)$};
\node[rotate=90] at (-0.5,2) (a4) {\bf string};
\node[rotate=90] at (6.5,2) (a5) {\bf relativistic};
\node[rotate=90] at (7,2) (a6) {\bf SSSSG};
\node at (1.5,3) (a6) {\bf magnons};
\node at (1.5,2.2) (b1) {$\langle a-1,0\rangle^{\times2}$};
\node at (4.5,1.5) (a7) {\bf solitons};
\node at (4.5,0.7) (b2) {$\langle0,a-1\rangle^{\times2}$};
\draw[->] (3,0.5) -- (3,3.2);
\node at (3,3.5) (a8) {$|p|$};
\begin{scope}[xshift=7.5cm]
\draw[-,very thick] (0,0) -- (6,0) -- (6,4) -- (0,4) -- (0,0);
\node at (6,-0.4) (a1) {$(g,\infty)$};
\node at (3,-0.4) (a3) {$(g,k)$};
\node[rotate=90] at (6.5,2) (a5) {\bf mirror string};
\node at (2.7,3) (a6) {\bf mirror magnons};
\node at (3.5,2.3) (a7) {$=$~~~ {\bf solitons}};
\node at (2.8,1.2) (b1) {$\langle 0,a-1\rangle^{\times2}$};\end{scope}
\draw[<->,very thick] (4,4.3) arc (110:70:8);
\node at (6.75,5.2) (c1) {double Wick rotation};
\end{tikzpicture}
\end{center}
In the theory described by the left-hand side of the diagram, the conventional stringy magnon theory is obtained in the limit $k\to\infty$ with fixed $g$ (see \cite{Beisert:2010jr} and references therein). The limit $g\to\infty$ with fixed $k$ of the same theory gives the relativistic semi-symmetric space sine-Gordon (SSSSG) theory, that is to say the Pohlmeyer reduced theory \cite{Grigoriev:2007bu,Mikhailov:2007xr,Hoare:2009fs,Hoare:2011fj,Hollowood:2011fq,Hoare:2011nd}. On this side of the diagram, at generic $(g,k)$ a state is either a soliton or magnon depending on whether the momentum is $|p|\lessgtr E$, respectively, an occurrence that is peculiar to a non-relativistic theory lacking Lorentz boost symmetry. Here, $a=1,2,\ldots$ labels the bound state which transforms in representation $\langle a-1,0\rangle^{\times2}$ for a magnon and $\langle 0,a-1\rangle^{\times2}$ for a soliton.

The right-hand side of the diagram describes the mirror theory obtained by a double Wick rotation. Such a transformation has no effect on the relativistic SSSSG theory but for generic $(g,k)$ the transformation is physically significant since the change in energy and momentum changes the bound state structure. In this case there is only one set of states, the mirror magnons or solitons which transform in representation $\langle0,a-1\rangle^{\times2}$. The stringy limit $k\to\infty$ with $g$ fixed of the mirror theory gives the theory that is needed to define the thermodynamic Bethe Ansatz of the string moving on $\text{AdS}_5\times S^5$~\cite{arXiv:0901.1417,Bombardelli:2009ns,arXiv:0902.4458,arXiv:0907.2647,arXiv:1012.3995}.

\section{Symmetries, Dispersion Relations and S-Matrix \label{sec2}}

The $q$ deformed theory admits a quantum group deformed version of the centrally-extended Lie super-algebra
\EQ{\label{symalg}
\mathfrak{psu}(2|2) \oplus \mathfrak{psu}(2|2)\ltimes\mathbb{R}^3 \ .
}
This algebra consists of two copies of the triply-extended super-algebra $\mh=\mpsu(2|2)\ltimes\mathbb{R}^3$ with the central extensions of the two factors identified. The fundamental states of the theory transform in the  
product of two copies of the short 4-dimensional representation $\langle0,0\rangle$ of $\mh$. We will denote this 16-dimensional product representation $\langle0,0\rangle^{\times2}$. The representation $\langle0,0\rangle$ has an associated module spanned by the basis $\{|\phi^a\rangle,|\psi^\alpha\rangle\}$, $a,\alpha=1,2$, where $|\phi^a\rangle$ are bosonic and $|\psi^\alpha\rangle$ are fermionic. The basic states of the theory are then the sixteen product states $|\phi^a\phi^b\rangle$, $|\phi^a\psi^\beta\rangle$, $|\psi^\alpha\phi^b\rangle$ and $|\psi^\alpha\psi^\beta\rangle$. Note that for a single representation $\langle0,0\rangle$, the bosonic and fermionic states are distinguished by the fermionic parity $(-1)^F|\phi^a\rangle=|\phi^a\rangle$ and $(-1)^F|\psi^\alpha\rangle=-|\psi^\alpha\rangle$, but the product states are actually invariant under interchanging the fermionic parity of $|\phi^a\rangle$ and $|\psi^\alpha\rangle$ (as is clear from \eqref{sws} below).

Each $\mpsu(2|2)$ has two $\msu(2)$ bosonic sub-algebras under which $|\phi^a\rangle$ transforms as a $(1,0)$ and $|\psi^\alpha\rangle$ as a $(0,1)$.\footnote{Here, we label the spin $j$ representations of $\msu(2)$ as $2j$.} Overall, the theory therefore has four $\msu(2)$ bosonic sub-algebras under which the representation $\langle0,0\rangle^{\times2}$ decomposes as
\EQ{\label{sws}
(1,0,1,0)\text{:}&\qquad|\phi^a\phi^b\rangle\qquad (-1)^F=1\ ,\\
(1,0,0,1)\text{:}&\qquad|\phi^a\psi^\beta\rangle\qquad (-1)^F=-1\ ,\\
(0,1,1,0)\text{:}&\qquad|\psi^\alpha\phi^b\rangle\qquad (-1)^F=-1\ ,\\
(0,1,0,1)\text{:}&\qquad|\psi^\alpha\psi^\beta\rangle\qquad (-1)^F=1\ ,
}
with fermionic parity as written.

Bound states are expected to transform in either the product of short representations $\langle a-1,0\rangle^{\times2}$ or $\langle 0,a-1\rangle^{\times2}$ each of dimension $16a^2$. Taken individually the two representations $\langle a-1,0\rangle$ and  $\langle 0,a-1\rangle$ are physically distinguished by the fermionic parity of the states. In fact the representations are sometimes called the ``symmetric" and ``anti-symmetric" representations, respectively, although their $\msu(2)\oplus\msu(2)$ are mirrors of each other:
\EQ{
\langle a-1,0\rangle&=(a,0)\oplus(a-1,1)\oplus(a-2,0)\ ,\\
\langle0,a-1\rangle&=(0,a)\oplus(1,a-1)\oplus(0,a-2)\ .
}
However, just as for the fundamental representation, once we take the product the two representations become isomorphic. 

Following \cite{Beisert:2011wq}, the fundamental particle states are labelled by two kinematic variables $x^\pm$ which are subject to a dispersion relation
\EQ{\label{p11}
q^{-1}\Big(x^++\frac1{x^+}\Big)-q\Big(x^-+\frac1{x^-}\Big)=(q-q^{-1})\Big(\xi+\frac1\xi\Big)\ ,
}
where
\EQ{\label{couplings}
\xi=-i\tilde g(q-q^{-1})\ ,\qquad
\tilde g^2=\frac{g^2}{1-g^2(q-q^{-1})^2}\ .
}
The deformation parameter $q$ is taken to be the complex phase\footnote{In the relativistic $g \rightarrow \infty$ limit, $k$ is identified with the coupling of the SSSSG or Pohlmeyer reduced theory \cite{Hoare:2011fj,Hollowood:2011fq}.}
\EQ{
q=\exp\Big[\frac{i\pi}k\Big]\ ,\qquad k\in\mathbb R>0\ .
}
With the choice above both $\tilde g$ and $\xi$ are real numbers with $\tilde g>0$ and $0\leq\xi\leq 1$. 
In \cite{Hoare:2011wr} it was shown that another useful way to parameterize the kinematic parameters $x^\pm$ is via the map $x(u)$ defined by
\EQ{\label{sqr}
x+\frac1x+\xi+\frac1\xi=\frac1\xi\left(\frac{\tilde g}{g}\right)^2 q^{-2iu}\ .
}
Note that $x(u)$ has two sheets. Given the map $x(u)$, we can then solve \eqref{p11} by taking
\EQ{
x^\pm=x\Big(u\pm \frac i2\Big)\ ,
}
which shows that the kinematic variables define a 4-sheeted branched cover of the $q^{-2iu}$ plane which defines the rapidity torus. The 4-fold cover of the $u$-plane has sheets
\EQ{\label{sheets} 
{\cal R}_{\pm2}\,:\qquad |x^+|<1\ ,\quad |x^-|<1\ , \\
{\cal R}_1\,:\qquad |x^+|<1\ ,\quad |x^-|>1\ , \\
{\cal R}_0\,:\qquad |x^+|>1\ ,\quad |x^-|>1\ , \\
{\cal R}_{-1}\,:\qquad |x^+|>1\ ,\quad |x^-|<1\ , 
}
where it is useful to think of the label as defined modulo 4 so that ${\cal R}_{-2}\equiv{\cal R}_2$. In some situations it is useful to introduce yet another coordinate $z$ which is the uniformized coordinate for the torus \cite{Hoare:2011wr}. 

The abstract kinematic variables $x^\pm$ encode the energy and momentum of the state. To make this concrete one can define the two quantities  \cite{Beisert:2011wq}
\EQ{\label{jww}
U^2=q^{-1}\frac{x^++\xi}{x^-+\xi}=q\frac{\frac1{x^-}+\xi}{\frac1{x^+}+\xi}\ ,\qquad
V^2=q^{-1}\frac{\xi x^++1}{\xi x^-+1}=q\frac{\frac\xi{x^-}+1}{\frac\xi{x^+}+1}\ ,
}
where the equalities follow from the dispersion relation \eqref{p11}. These quantities determine the three central charges $(C,P,K)$ of the symmetry algebra \eqref{symalg} by
\EQ{\label{cch}
q^{2C}=V^2\ ,\qquad P=g\alpha(1-U^2V^2)\ ,\qquad K=g\alpha^{-1}(V^{-2}-U^{-2})\ ,
}
where $\alpha$ is a free normalization constant. The central charges are related via
\EQ{\label{sht}
[C]_q^2-PK=[1/2]_q^2\, ,
}
where $[x]_q=(q^x-q^{-x})/(q-q^{-1})$. This is precisely the shortening condition for the atypical, or short, fundamental 4-dimensional representation of $U_q(\mh)$ labelled $\langle0,0\rangle$. What is particularly noteworthy is that the Hopf algebra structure ensures that the group-like elements $U$ and $V$ have a trivial co-product on a multi-particle state, e.g.~on a two particle state
\EQ{\label{cop}
\Delta(U)=U\otimes U\ ,\qquad \Delta(V)=V\otimes V\ .
}
The dispersion relation \eqref{p11} can be re-written in terms of $U$ and $V$ as
\EQ{\label{dsr}
V^2+V^{-2}-g^2(q-q^{-1})^2(V^2+V^{-2}-U^2-U^{-2})=q+q^{-1}\ ,
}
or equivalently (see appendix \ref{appa})
\EQ{\label{dit2}
\xi^{-1}\left(V^2 + \frac1{V^{2}} - q - \frac1q\right) = \xi \left(U^2 + \frac1{U^2} - q - \frac1q\right)\ .}

\noindent{\bf Magnon kinematics}

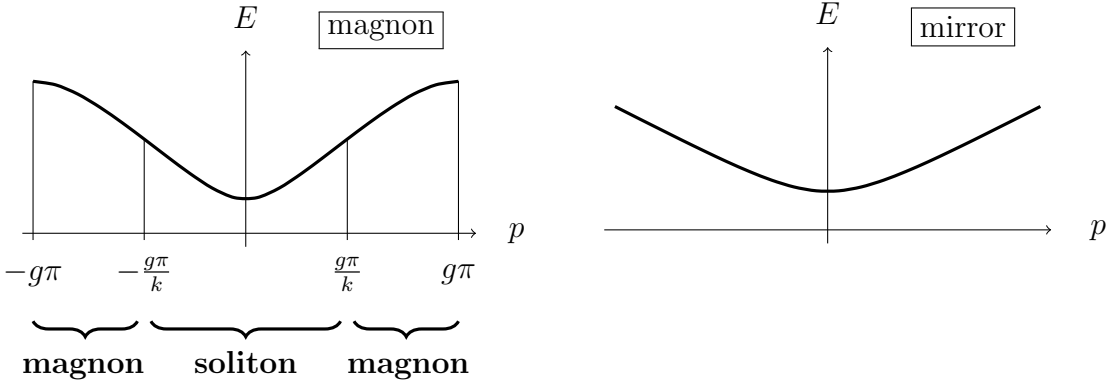
\begin{figure}[ht]
\begin{minipage}[t]{.49\textwidth}
\begin{center}
\begin{tikzpicture}[scale=0.9]
\draw[->] (-3.3,0) -- (3.4,0);
\draw[->] (0,-0.2) -- (0,2.7);
\draw[-] (-3.142,2.25) -- (-3.142,-0.1);
\draw[-] (3.142,2.25) -- (3.142,-0.1);
\draw[-] (-1.5,1.4) -- (-1.5,-0.1);
\draw[-] (1.5,1.4) -- (1.5,-0.1);
\node at  (1.5,-0.6) (c1) {$\frac{g\pi}k$};
\node at  (-1.5,-0.6) (c2) {$-\frac{g\pi}k$};
\node at (0,-1.9) (b1) {\bf soliton};
\node at (2.4,-2) (b2) {\bf magnon};
\node at (-2.4,-2) (b3) {\bf magnon};
\node at (3.142,-0.6) (a1) {$g\pi$};
\node at (-3.142,-0.6) (a2) {$-g\pi$};
\node at (4,0) (a3) {$p$};
\node at (0,3.2) (a4) {$E$};
\draw[very thick] plot[smooth] coordinates {(-3.142, 2.246)  (-2.811, 2.195)  (-2.480, 2.057)  (-2.150, 1.860)  (-1.819, 1.630)  (-1.488, 1.383)  (-1.157, 1.130)  (-0.8267, 0.8854)  (-0.4960, 0.6687)  (-0.1653, 0.5261)  (0.1653, 0.5261)  (0.4960, 0.6687)  (0.8267, 0.8854)  (1.157, 1.130)  (1.488, 1.383)  (1.819, 1.630) (2.150, 1.860)  (2.480, 2.057)  (2.811, 2.195)  (3.142, 2.246)};
\node at (2,3) (b1) {$\boxed{\text{magnon}}$};
\draw[decoration={brace,amplitude=0.5em},decorate,very thick] (1.4,-1.3) -- (-1.4,-1.3);
\draw[decoration={brace,amplitude=0.5em},decorate,very thick] (3.142,-1.3) -- (1.6,-1.3);
\draw[decoration={brace,amplitude=0.5em},decorate,very thick] (-1.6,-1.3) -- (-3.142,-1.3);
\end{tikzpicture}
\end{center}
\end{minipage}
\hfill
\begin{minipage}[t]{.49\textwidth}
\begin{center}
\begin{tikzpicture}[scale=0.9]
\draw[->] (-3.3,0) -- (3.3,0);
\draw[->] (0,-0.2) -- (0,2.7);
\node at (4,0) (a3) {$p$};
\node at (0,3.2) (a4) {$E$};
\node at (0,-2.2) (q1) {};
\draw[very thick] plot[smooth] coordinates {(-3.142, 1.824)  (-2.811, 1.656)  (-2.480, 1.490)  (-2.150,  1.325)  (-1.819, 1.164)  (-1.488, 1.008)  (-1.157,  0.8632)  (-0.8267, 0.7347)  (-0.4960, 0.6341)  (-0.1653, 0.5771)  (0.1653, 0.5771)  (0.4960, 0.6341)  (0.8267, 0.7347)  (1.157, 0.8632)  (1.488, 1.008)  (1.819, 1.164)  (2.150, 1.325)  (2.480, 1.490)  (2.811, 1.656)  (3.142, 1.824)};
\node at (2,3) (b1) {$\boxed{\text{mirror}}$};
\end{tikzpicture}
\end{center}
\end{minipage}
\caption{\small Dispersion relation for ``magnon kinematics" (left)---in this case the momentum is restricted to lie in the interval $|p|\leq g\pi$, and ``mirror kinematics" (right)---in this case the momentum takes values along the real line. For the case of magnon kinematics we have shown the two distinct branches labelled ``soliton" and ``magnon" corresponding to $|p|\lessgtr E$, that is $|p|\lessgtr\pi g/k$, respectively.}
\label{f1}
\end{figure}

The fact that $U$ and $V$ have a trivial co-product suggest that they should be related in a simple way to energy and momentum. To start with let us take magnon kinematics identified by the behaviour in the string theory limit, that is $k\to\infty$ with $g$ fixed. In this case, the known behaviour suggests that physical states are subject to the reality condition $x^{\pm*}=x^\mp$. This implies that $U^*=U^{-1}$ and $V^*=V^{-1}$. The relation with the energy and momentum (up to some overall re-scaling) is
\EQ{\label{uvr1}
U=e^{ip/2g}\ ,\qquad V=e^{iE/2g}\ .
}
To define a one-to-one map between $p,E$ and the central charges $(C,P,K)$ given by~\eqref{cch}, we restrict the arguments of $U^2$ and $V^2$ to lie in the range $[-\pi,\pi]$, so that the momentum and the energy lie in the region $|p|, |E|\leq \pi g$. It is worth noticing that the coproduct~\eqref{cop} ensures  the conservation of energy and momentum only modulo $2\pi g$. This suggests that, for generic finite values of $(g,k)$, the interpolating theory with magnon kinematics could exhibit umklapp type processes, an interesting possibility that will be investigated elsewhere.\footnote{While the periodic dependence of the dispersion relation on $E$ and $p$ is not an issue for finding the solutions to the bound state equations (the subject we address in this paper), a more general question is how to define the energy or momentum of a multi particle state. In particular, defining energy and momentum modulo $2\pi g$ implies that a tensor product of two single particle states with physical values of energy and positive momentum can have a total energy or momentum which is negative. While for momentum this is well-understood in the context of lattice models (giving rise to umklapp type processes), for energy it is not clear how this should be interpreted physically.} Substituting in for $U$ and $V$ the dispersion relation \eqref{dsr} becomes
\EQ{\label{dismag}
\cos(E/g)-\xi^2\cos(p/g)=(1-\xi^2)\cos(\pi/k)\ .
}
Taking $|p|\leq \pi g$, physical values of the energy should correspond to positive solutions of the dispersion relation \eqref{dismag}. This places physical values of the kinematic variables on the sheet ${\cal R}_0$ of the rapidity torus. The left-hand side of Fig.~\ref{f1} shows the dispersion relation. 

In the string limit we have $p$ of order $k^0$, $E$ of order $k^{-1}$ and the dispersion relation becomes the familiar \cite{Beisert:2004hm}
\EQ{
E^2=\Big(\frac{\pi g}{2k}\Big)^2\Big(1+16g^2\sin^2\frac{g\,p}2\Big)\ ,
}
up to a simple re-scaling of $E$ and $p$. In the relativistic limit, that is $g\to\infty$ with fixed $k$, the dispersion relation \eqref{dsr} assumes the normal relativistic form
\EQ{
E^2-p^2=\left(2\cos\frac\pi{2k}\right)^{-2}\ .
}

It is interesting to investigate the relation between the physical momentum and the generalized rapidity variable $u$ \eqref{sqr}. In fact one finds that the reality condition $x^{\pm*}=x^\mp$ implies that $q^{-2iu}$ is real but there are two distinct branches depending on whether $q^{-2iu}$ is real and positive or negative. Using the relations
\EQ{\label{xpm}
x^+&=\xi\cdot\frac{1-qU^2}{qV^{-2}-1}=\frac1\xi\cdot\frac{q V^2-1}{1-qU^{-2}}\ ,\\
x^-&=\xi\cdot\frac{U^{-2}-q}{q-V^2}=\frac1\xi\cdot\frac{q-V^{-2}}{U^2-q}\ ,
}
where the equalities are guaranteed by the dispersion relation \eqref{dsr}, we can correlate the momentum $p$ with the coordinate $u$ as
\EQ{\label{claim}
\text{soliton branch:}\qquad& |p|<E\ ,\qquad|p|< \frac{\pi g}k\ ,\qquad q^{-2iu}\in{\mathbb R}<0\ ,\\
\text{magnon branch:}\qquad&|p|>E\ ,\qquad \frac{\pi g}k< |p| \ ,\qquad q^{-2iu}\in{\mathbb R}>0\ .
}
Note that the cross-over is precisely when $|p|=E$. An explicit proof of this claim follows from the relation is provided by the simple on-shell relation \eqref{rr}
\EQ{
q^{-2iu} = \frac{2\sin^2(\frac{E+p}{2g})}{(1-\xi^2)(\cos \frac{\pi}{k}-\cos \frac{p}{g})}
         = \frac{\sin(\frac{p+E}{2g})}{\sin(\frac{p-E}{2g})}\ .
}
proved in appendix \ref{appa}. The names soliton and magnon in \eqref{claim} refer to the fact that in the string limit $k\to\infty$ only the magnon branch remains, while in the relativistic limit $g \to \infty$ only the soliton branch remains. 

It is worth pausing to appreciate what is going on here. In a relativistic theory, Lorentz transformations can be used to boost the momentum of a state. The intrinsic properties of the state, however, do not change. A remarkable finding of our analysis is that this simple intuition must be abandoned in the non-relativistic interpolating theory: the nature of the state depends on its momentum.

\noindent{\bf Mirror kinematics}

There is an associated theory, the so-called mirror theory, obtained by a double Wick rotation $p\to -iE$ and $E\to -ip$~\cite{arXiv:0710.1568}. In particular, the S-matrix of the mirror theory is just the S-matrix of the original theory with this transformation on the kinematic variables. In the relativistic limit, the mirror theory is identical to the original but in the more general non-relativistic theory, for generic $(g,k)$ including the string theory limit, this is no longer true. With mirror kinematics we have
\EQ{\label{uvr2}
U=e^{E/2g}\ ,\qquad V=e^{p/2g}
}
and physical states satisfy a different reality condition $x^{+*}=1/x^-$ which ensures $U^*=U$ and $V^*=V$. Notice that, in contrast to the case of magnon kinematics, the coproduct~\eqref{cop} is consistent now with the usual conservation of energy and momentum. In the mirror case, the dispersion relation becomes 
\EQ{\label{dismirr}
\cosh(p/g)-\xi^2\cosh(E/g)=(1-\xi^2)\cos(\pi/k)\ ,
}
where the momentum $p$ is valued along the real line. The physical values of the energy are the positive solutions of~\eqref{dismirr}, and on the right-hand side of Fig.~\ref{f1} we illustrate this dispersion relation. The corresponding kinematic variables lie on the sheet ${\cal R}_{-1}$ of the rapidity torus.
 
The relativistic dispersion relation results from taking the limit $g\to\infty$, in which case 
\EQ{\label{rlm}
E^2-p^2=\left(2\cos\frac\pi{2k}\right)^{-2}\ .
}
This is identical to the same limit of the magnon theory as expected since the double Wick rotation is a symmetry of the relativistic theory. In the string limit $k\to\infty$, the dispersion relations becomes
\EQ{
E=2g\,\text{arcsinh}\,\Big[ \frac1{4g}\sqrt{1+\Big(\frac{2kgp}\pi\Big)^2}\Big]\ ,
}
which with a simple re-scaling of $E$ and $p$ is the known dispersion relation of the mirror theory \cite{arXiv:0710.1568}.

Concerning the relationship between the physical momentum and the generalized rapidity variable $u$, in the mirror case the reality condition $x^{+*}=1/x^-$ also implies that $q^{-2iu}$ is real. But, compared to the case of magnon kinematics, $q^{-2iu}$ turns out to be always positive, as can be easily checked using eq.~\eqref{rrpre}.

We will not write down the complete S-matrix. However, it is worth writing down two of its elements that play an important r\^ole. The first fact to bear in mind is that when written in terms of the $x_1^\pm$ and $x_2^\pm$ variables the S-matrix for the magnon and mirror theories are identical. It is only the way that $x_1^\pm$ and $x_2^\pm$ are expressed in terms of the energy and momentum that changes. The S-matrix element of the so-called ``$\msu(2)$ sector" corresponding to the particular element $|\phi^a\phi^a;z_1\rangle\otimes\ket{\phi^a\phi^a;z_2}\to|\phi^a\phi^a;z_2\rangle\otimes\ket{\phi^a\phi^a;z_1}$ takes the form
\EQ{\label{xpp}
S_{\msu(2)}(z_1,z_2)=\frac1{\sigma(z_1,z_2)^2}\cdot\frac{x_1^+x_2^-}{x_1^-x_2^+}\cdot\frac{x_1^--x_2^+}{x_1^+-x_2^-}\cdot \frac{1-\frac1{x_1^-x_2^+}}{1-\frac1{x_1^+x_2^-}}\ .
}
In the above, we have written the energy and momentum of the states in terms of the uniformized coordinate $z$ \cite{Hoare:2011wr}. The other S-matrix element of interest is the one of the ``$\msl(2)$ sector"\footnote{This terminology of $\msu(2)$ and $\msl(2)$ sectors is not completely appropriate since all states transform in representations of the compact bosonic algebra $\msu(2)^{\oplus4}$. The origin refers to the fact that, in the context of $\text{AdS}_5\times S^5$, the pairs of $\msu(2)$'s under which $|\phi^a\phi^b\rangle$ transforms as $(2,2)$ form a subalgebra of the algebra of the compact $\text{SO(6)}$ arising from the $S^5\simeq\text{SO}(6)/\text{SO}(5)$  part of the geometry whilst  the pair of $\msu(2)$'s under which $|\psi^\alpha\psi^\beta\rangle$ transforms as $(2,2)$ are a subalgebra of the algebra of the non-compact group $\text{SO(2,4)}$ arising from the $\text{AdS}_5\simeq\text{SO}(2,4)/\text{SO}(1,4)$  part of the geometry.} that corresponds to the particular element $|\psi^\alpha\psi^\alpha;z_1\rangle\otimes\ket{\psi^\alpha\psi^\alpha;z_2}\to|\psi^\alpha\psi^\alpha;z_2\rangle\otimes\ket{\psi^\alpha\psi^\alpha;z_1}$, which takes the form 
\EQ{\label{xpp2}
S_{\msl(2)}(z_1,z_2)=\frac1{\sigma(z_1,z_2)^2}\cdot\frac{x_1^+-x_2^-}{x_1^--x_2^+}\cdot \frac{1-\frac1{x_1^-x_2^+}}{1-\frac1{x_1^+x_2^-}}\ .
}

In a relativistic theory, the S-matrix is valued on an infinite but unbranched cover of the relative rapidity cylinder. This is simply the relative rapidity complex plane $\theta=\theta_1-\theta_2$. Physical values correspond to real values of $\theta$. The situation in the non-relativistic interpolating theory is much more complex. The S-matrix is a function separately of the rapidity of each of the incoming particles since there is no relativistic invariance or notion of a relative rapidity. Then, for each of the rapidities the S-matrix is actually a function on an infinite but now branched cover of the rapidity torus (the latter with uniformized coordinate $z$). This cover cannot be viewed as the complex $z$ plane, however, it is useful to think of it as extending the sheets ${\cal R}_n$, with $n=-1,0,1,2$, to $n\in{\mathbb Z}$ where now the sheet ${\cal R}_{-2}$ is not identified with ${\cal R}_2$. In this description, we label the sheets for both particles together as ${\cal R}_{m,n}\equiv{\cal R}_m^{(1)}\times{\cal R}_n^{(2)}$. For instance, physical values of the rapidity of the incoming particles lie on sheet ${\cal R}_{0,0}$ for magnon kinematics, and sheet ${\cal R}_{-1,-1}$ for mirror kinematics. But even this description is only partial since there are a plethora of additional cuts that lead to further sheets. The dressing phase $\sigma(z_1,z_2)$ is regular in the region ${\cal R}_{0,0}$ and as a consequence the $\msu(2)$ sector S-matrix \eqref{xpp} has a simple pole at $x_1^+=x_2^-$ while the $\msl(2)$ sector S-matrix \eqref{xpp2} has a simple pole at $x_1^-=x_2^+$. In the next section we shall investigate which of these poles lies on the analogue of the physical sheet and so is a candidate for a bound state pole. The analytic continuation of the dressing phase to other sheets is complicated but can be deduced from the formulae in~\cite{Hoare:2011wr}.

\section{Bound states}

In a relativistic theory, two-to-two body particle scattering can be described in terms of the familiar $s=(p_1+p_2)^2$, $t=(p_1-p_3)^2$ and $u=(p_1-p_4)^2$ Mandelstam variables. In a 1+1 dimensional integrable theory the individual momenta are conserved, $p_3=p_2$ and $p_4=p_1$, so that $u=0$ and $t=2(m_1^2+m_2^2) -s$. Considering equal mass scattering ($m_1 = m_2 = m$) and introducing the rapidity $p_i=m(\cosh\theta_i,\sinh\theta_i)$ we have
\EQ{
s=4m^2\cosh^2\frac\theta2\ ,
}
where $\theta=\theta_1-\theta_2$. S-matrix theory involves how the S-matrix behaves in the complex $s$-plane. In fact the S-matrix is defined on an infinite cover of the complex plane. One starts on the {\it physical sheet\/} which has branch points at $s=4m^2$, the 2-particle threshold, and $s=0$. The cuts are usually taken to lie along the positive and negative real axis, respectively. In the complex rapidity plane the physical sheet becomes the {\it physical strip\/}  $0\leq\IM\theta\leq\pi$. In an integrable theory the infinite set of sheets of $s$ cover the whole complex rapidity plane. In this sense, the rapidity is a uniformizing coordinate for the S-matrix. Physical values correspond to $s=s_0+i0^+$ with $s_0$ real and $\geq4m^2$. This corresponds to $\theta$ being real and positive. The existence of a stable bound state of mass $M$ is signalled by a simple pole at $s=M^2$ lying along the real axis between $s=0$ and $s=4m^2$,\footnote{Other poles can correspond to bound states in the crossed channel or anomalous thresholds.} where the upper bound here is the 2-particle threshold. In terms of the rapidity, this corresponds to imaginary values of $\theta$ between 0 and $\pi$. In terms of the spatial momentum these bound states occur at complex values $p_1=\tilde p+ir$ and $p_2=\tilde p-ir$. Note that the fact that the energy of the bound state is real requires that the real parts of the momenta are equal. The imaginary part of the momentum $r$ is determined by the mass-shell condition for the bound state $s=M^2$. For the case of incoming particles with equal mass $m$ we have
\EQ{\label{bsc}
r^2=\Big(\frac{4m^2}{M^2}-1\Big)\Big(\tilde p^2+\frac{M^2}4\Big)\ .
}

The problem that we face is that it is not obvious how to generalize these aspects of S-matrix theory to the non-relativistic context and in particular how to identify when a pole in the S-matrix corresponds to a bound state, an anomalous threshold, or neither. The conditions under which a bound state can form can most easily be stated in terms of the variables $x_i^\pm$. From the representation theory of the symmetry algebra, bound states can occur when $x_1^+=x_2^-$, corresponding to the product of the short representations $\langle1,0\rangle^{\times2}$, or when $x_1^-=x_2^+$, corresponding to the product of the short representations $\langle0,1\rangle^{\times2}$. More generally we expect that the theory contains states in representations $\langle a-1,0\rangle^{\times2}$ and/or $\langle 0,a-1\rangle^{\times 2}$, with $a=1,2,\ldots$. The general dispersion relation for these states takes the form\footnote{Using the dispersion relation in terms of $x^\pm$ in \cite{deLeeuw:2011jr,Hoare:2011wr}; namely,
\EQ{\label{bsdr}
q^{-a}\Big(x^++\frac1{x^+}\Big)-q^a\Big(x^-+\frac1{x^-}\Big)=(q^a-q^{-a})\Big(\xi+\frac1\xi\Big)
}
and
\EQ{\label{bsuv}
U^2=q^{-a}\frac{x^++\xi}{x^-+\xi}=q^a\frac{\frac1{x^-}+\xi}{\frac1{x^+}+\xi}\ ,\qquad
V^2=q^{-a}\frac{\xi x^++1}{\xi x^-+1}=q^a\frac{\frac\xi{x^-}+1}{\frac\xi{x^+}+1}\ .
}
With magnon and mirror kinematics, the physical values of the kinematic variables of these states also lie on the sheets ${\cal R}_0$ and ${\cal R}_{-1}$, respectively.}
\EQ{\label{dsr2}
V^2+V^{-2}-g^2(q-q^{-1})^2(V^2+V^{-2}-U^2-U^{-2})=q^a+q^{-a}\ ,
}
with $(U,V)$ related to the energy and momentum as in \eqref{uvr1} or \eqref{uvr2}. In particular, in the relativistic limit, one finds
\EQ{\label{rlm2}
E^2-p^2=\Big(\frac{q^{a/2}-q^{-a/2}}{q-q^{-1}}\Big)^{2}=\left[a/2\right]_q^2\ .
}

With magnon kinematics, the situation we found in section \ref{sec2} for the fundamental particle generalizes in an obviously way (see eq.~\eqref{rrbs}). Once again there are two branches, a soliton branch with $|p|<a\pi g/k$ and a magnon branch with $|p|>a\pi g/k$. In a moment we will argue that the bound states in each branch are actually different since they occur at $x_1^-=x_2^+$ on the soliton branch and at $x_1^+=x_2^-$ on the magnon branch. In particular, this means that the bound states transform in representations $\langle0,a-1\rangle^{\times2}$ for $|p|<a\pi g/k$ and $\langle a-1,0\rangle^{\times2}$ for $|p|>a\pi g/k$. We will argue that at the special point $|p|=a\pi g/k$ the states become un-bound. This behaviour, as we have already described for the fundamental particle, clearly could not occur in a relativistic theory due to the principle of special relativity. 

In the non-relativistic theory, we can find the generalization of the bound state condition \eqref{bsc} by imposing the dispersion relation \eqref{dsr2} for the bound state $a=2$ on the incoming energy and momenta $U=U_1U_2$ and $V=V_1V_2$, that is $E=E_1+E_2$ and $p=p_1+p_2$ up to the comments made after eq.~\eqref{uvr1} for magnon kinematics. In the non-relativistic theory we do not know a priori the analogue of the physical sheet/strip so it is not entirely obvious how to identify which simple poles of the S-matrix are due to bound states in the direct channel. However, we can proceed in a way that is as close as possible to the relativistic case. The idea is to take an analytic continuation of the incoming momenta  away from physical values along the path $p_1=\tilde p_1+ir$ and $p_2=\tilde p_2-ir$ with $r$ increasing from 0, keeping the overall energy $E_1+E_2$ real, until the total energy and momentum hit the on-shell values for the bound state. In a relativistic theory, the condition that the total energy is real, with equal masses for the incoming particles, requires $\tilde p_1=\tilde p_2$ and we shall find the same condition in the non-relativistic case for small enough $\tilde p_1$. As we discuss below, a necessary condition for the formation of a bound state is $r>0$. In the relativistic situation the analytic continuation described above covers the region where $s\in{\mathbb R}$ and $0<s <4m^2$ lying on the physical strip, precisely the region where we expect to find bound state poles. We will assume that bound state poles must be found in the same analytically continued region in the $q$-deformed non-relativistic theory.

The positivity of $r$ follows thinking about scattering in one of the scalar sectors, e.g.~the $\msu(2)$ sector with S-matrix \eqref{xpp}. The argument, taken from \cite{Dorey:2007xn}, is a simple application of quantum mechanical scattering theory (for example, see~\cite{Mussardo:2010} appendix 17B). Labelling the S-matrix by the momenta of the incoming excitations, the wavefunction for the 2 states in the asymptotic regime $x_1\ll x_2$\,\footnote{Note that here $x_{1,2}$ are the spatial coordinates of the two particles being scattered and are not related to $x_{1,2}^\pm$.} can be expressed as
\EQ{
\Psi(x_1,x_2)=e^{ip_1x_1+ip_2x_2}+S(p_1,p_2)e^{ip_1x_2+ip_2x_1}\ .
}
Assuming that $v_1>v_2$,\footnote{The velocities are given by $v=\partial E/\partial p$ up to an overall positive normalization.} the first term represents the incoming two-particle wave and the second term the reflected wave with permuted momenta. We then analytically continue the momenta $p_1=\tilde p_1+ir$ and $p_2=\tilde p_2-ir$. The incoming wavefunction then behaves as $e^{-r(x_1-x_2)}$ and the reflected wavefunction as $e^{r(x_1-x_2)}$ in the region $x_1\ll x_2$. In order that we can give a probabilistic interpretation to the formation of a bound state we require that the reflected wavefunction is small relative to the incoming wavefunction in the region $x_1\ll x_2$ which means that a bound state can only form when the imaginary part of the momentum of the incoming particle from the left is positive: $r>0$. 

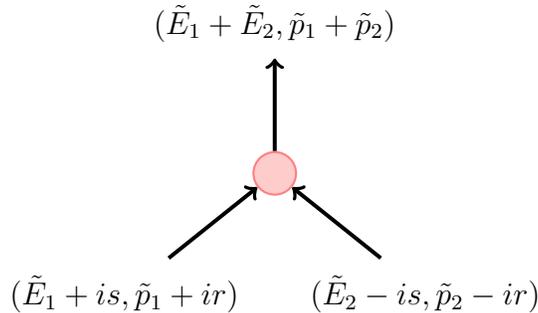
\begin{figure}[ht]
\begin{center}
\begin{tikzpicture} [line width=1.5pt,inner sep=2mm,
place/.style={circle,draw=blue!50,fill=blue!20,thick},proj/.style={circle,draw=red!50,fill=red!20,thick}]
\node at (2,2) [proj] (p1) {};
\node at (0,0.4) (i1) {$(\tilde E_1+is,\tilde p_1+ir)$};
\node at (4,0.4) (i2)  {$(\tilde E_2-is,\tilde p_2-ir)$};
\node at (2,4) (i3) {$(\tilde E_1+\tilde E_2,\tilde p_1+\tilde p_2)$};
\draw[->]  (i1) -- (p1);
\draw[->]  (i2) -- (p1);
\draw[<-]  (i3) -- (p1);
\end{tikzpicture}
\caption{\small The kinematic conditions required to form a bound state. All particles are on shell but the incoming particles have momenta which are analytically continued $p_1=\tilde p_1+ir$ and $p_2=\tilde p_2-ir$. Additional conditions to form a bound state are that $r>0$, $v_1\geq v_2$ and the bound state has physical (real) momentum and energy. The on-shell requirement implies that either $x_1^+=x_2^-$ or $x_1^-=x_2^+$. In the case of magnon kinematics,  $\tilde E_1+\tilde E_2$ and $\tilde p_1+\tilde p_2$ are restricted to be $>0$ and $<\pi g$, and $>-\pi g$ and $<\pi g$, respectively, and have to be understood modulo $2\pi g$.}
\label{f2}
\end{center}
\end{figure}

In the non-relativistic theory there is no a priori reason why the bound state should form with $\tilde p_1=\tilde p_2$. As stated above, the actual condition is determined by requiring that the energy of the bound state is real and positive. In the following analysis we find that this requires $\tilde p_{1,2}$ to have the same sign. Furthermore, restricting the discussion to $\tilde p_{1,2}>0$,\footnote{The analysis is symmetric with respect to $\tilde p_{1,2} \rightarrow - \tilde p_{2,1}$. Care needs to be taken when $\tilde p_1 \neq \tilde p_2$, in which case their r\^oles are interchanged on flipping signs. This is required to preserve $v_1 > v_2$.} we find that for $\tilde p_{1,2}$ both less than a critical value $\tilde p_0$ the reality of the total energy does imply that $\tilde p_1=\tilde p_2$ as in a relativistic theory. However, there exists another branch of solutions with $\tilde p_2> \tilde p_0$, i.e.~above the threshold, such that $\tilde p_1$ and $\tilde p_2$ are not equal and indeed $\tilde p_1<\tilde p_0$.\,\footnote{This is precisely what Arutyunov and Frolov found in \cite{arXiv:0710.1568} for the string theory limit.} Notice that there are other solutions with $\tilde p_1$ and $\tilde p_2$ interchanged but in our convention particle 1 is the one coming in from the left and so we require $v_1\geq v_2$ for consistency, where we identify the velocity with the real part of $\partial E/\partial p$.  This picks out the solution with $\tilde p_2>\tilde p_0>\tilde p_1$. The consistency conditions are illustrated in Fig.~\ref{f2}.

In order to solve the bound state conditions it is useful to notice that we can solve for $U$ and $V$ in terms of either $x^+$ or $x^-$. Using \eqref{xpm} we find the two equations
\EQ{\label{a11}
\Big(\frac1{x^\pm}+\xi\Big)\, U^{\pm2}+(x^\pm+\xi)\, U^{\mp2} = q^{\mp 1}\Big(x^\pm + \frac1{x^{\pm}} + \xi + \frac1\xi\Big) + q^{\pm 1}\Big(\xi - \frac1\xi\Big)\ ,
}
which can be solved for either $U=U(x^+)$ or $U=U(x^-)$ and then 
\EQ{\label{kqq}
V^2= \xi\big(q^{\mp1}-U^{-2}\big)(x^\pm)^{\pm1}+q^{\mp1}
}

\begin{figure}[ht]
\begin{center}
\begin{tikzpicture}[scale=1.2]
\draw[-] (0,0) -- (0,6) -- (6,6) -- (6,0) -- (0,0);
\draw[very thick] (1,0) .. controls (1,1) and (1.5,2) .. (2.5,2);
\draw[very thick,densely dashed] (5.5,0) .. controls (5.5,1) and (3.5,2) .. (2.5,2);
\draw[very thick,red] (2,0) .. controls (2.3,0.3) and (2.5,1) .. (2.5,2);
\draw[very thick,blue] (2,6) .. controls (2,5) and (2.5,3) .. (2.5,2);
\filldraw[black] (2.5,2) circle (1pt);
\draw[-] (0,2) -- (0.1,2);
\node at (-0.4,2) (z1) {$\tilde p_0$};
\node at (3,-0.7) (a1) {$r$};
\node at (-0.8,3) (a2) {$\tilde p$};
\node[blue] at (3.3,3.3) (a3) {$\tilde p_2$};
\node[red] at (3,0.5) (a4) {$\tilde p_1$};
\node at (1,2.5) (d1) {$|x_1^-|=1$};
\node[blue] at (1,4) (d2) {$x_1^-\in{\mathbb R}$};
\draw[->] (d1) -- (1.5,1.7);
\draw[->,blue] (d2) -- (2.1,4);
\node at (4.5,5.7) (b1) {mirror kinematics};
\node at (-0.4,0) (c3) {0};
\draw[->,blue] (a3) -- (2.5,3.3);
\draw[->,red] (a4) -- (2.4,0.5);
\end{tikzpicture}
\end{center}
\caption{\small Schematic plot (values exaggerated for clarity) of the bound-state momenta of the 2 particles $p_1=\tilde p_1+ir$ and $p_2=\tilde p_2-ir$ for mirror kinematics. The solid black and dashed part of the curve corresponds to the branch with $x_1^-=x_2^+$ and $|x_1^-|=1$ on which $\tilde p_1=\tilde p_2\leq\tilde p_0$. The solid blue/red part of the curve corresponds to the branch also with $x_1^-=x_2^+$ but now $x_1^-\in{\mathbb R}$, $\tilde p_2\geq p_0$ (blue) and $\tilde p_1\leq\tilde p_0$ (red).}
\label{f3}
\end{figure}
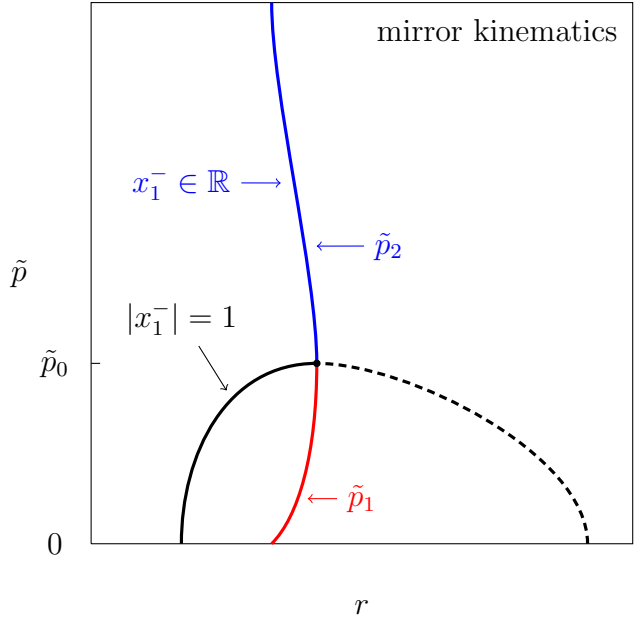

Bound states can form either when $x_1^- = x_2^+$ or $x_1^+ = x_2^-$. Working with the former and substituting these expressions into $U_1U_2$ and $V_1V_2$ it is clear that the parameters $x_{\!B}^\pm$ labelling the bound state are equal to $(x_1^+, x_2^-)$. As described above, in the fusion procedure the energy and momentum of the bound state should be a real---the reality conditions for the bound state (for both magnon and mirror kinematics) imply the following relation\footnote{This relation encodes both the magnon, $x_{\!B}^{+*} = x_{\!B}^-$, and mirror, $x_{\!B}^{+*} = 1/x_{\!B}^-$, reality conditions.}
\EQ{
\left(x_1^+ + \frac{1}{x_1^+}\right)^* = x_2^- + \frac{1}{x_2^-} \ . 
}
This condition along with the dispersion relations for $x_1^\pm$ and $x_2^\pm$ \eqref{p11} imply that
\EQ{
x_1^- + \frac{1}{x_1^-} \in \R \qquad (x_1^- = x_2^+)
}
is a necessary requirement for the formation of a bound state with real energy and momentum. Therefore, in the following discussion we can restrict the search for solutions of the bound state conditions to two cases: $x_1^- \in \R$ and $|x_1^-| = 1$. A similar argument works for the bound states forming when $x_1^+ = x_2^-$; that is either $x_1^+ \in \R$ or $|x_1^+| = 1$.

In addition, in order to be physical, the energy of the bound state has to be positive. In general, the dispersion relations~\eqref{p11} and~\eqref{bsdr} have two different solutions corresponding to positive and negative energy. An efficient way to select the physical one is to require that the kinematic variables of the bound state $x_B^\pm$ lie on the correct sheet of the rapidity torus. Namely, for magnon kinematics $x_B^\pm \in {\cal R}_0$ while for mirror kinematics  $x_B^\pm \in {\cal R}_{-1}$.

\noindent{\bf Mirror kinematics}

The equations above, \eqref{a11} and \eqref{kqq}, can be used to solve for the kinematic conditions at a bound state pole. Firstly, we consider the simpler case of mirror kinematics for which we have
\EQ{
&V_1=e^{(\tilde p_1+ir)/2g}\ ,\qquad V_2=e^{(\tilde p_2-ir)/2g}\ ,\\
&U_1=e^{(\tilde E_1+is)/2g}\ ,\qquad U_2=e^{(\tilde E_2-is)/2g}\ ,
}
a parameterization which ensures that the bound state has real momentum and energy. One then finds that the bound state condition $x_1^+=x_2^-$ has no solutions which satisfy the physical requirements. On the other hand for $x_1^-=x_2^+$, which corresponds to the bound state $\langle0,1\rangle^{\times2}$, we find that there are two distinct branches of solutions. On the first branch $|x_1^-|=1$, $\tilde p_1=\tilde p_2$ and $\tilde E_1=\tilde E_2$ whilst on the second $x_1^-\in{\mathbb R}$ and
\EQ{
e^{(\tilde p_1-\tilde p_2)/g}=\frac{\frac\xi{x_1^{-}}+1}{\xi x_1^-+1}\ ,\qquad e^{(\tilde E_1-\tilde E_2)/g}=\frac{\xi+x_1^-}{\xi+\frac1{x_1^-}}\ .
}
It is now a simple matter to map out the bound state momenta and energies using $V_1(x_1^-)$, $U_1(x_1^-)$, $V_2(x_2^+)$ and $U_2(x_2^+)$. Once the physical conditions are imposed, we find the solutions that are illustrated in Fig.~\ref{f3}. For $\tilde p_{1} = \tilde p_2 <\tilde p_0$, where the critical value $\tilde p_0$ corresponds to the point where the two branches meet at $x_1^-=1$, the solution corresponds to the first branch with $|x_1^-|=1$. On this branch $\tilde p_1=\tilde p_2$ and $\tilde E_1=\tilde E_2$. For a given $\tilde p_1$ there are actually two solutions for $r$, shown by the solid and dashed lines, and both are potential bound state poles. We discuss this in more detail below. This branch lies on the intersection of boundaries of the sheets ${\cal R}_{-1,-1}$ and  ${\cal R}_{0,-2}$. On the second branch for which $x_1^-\in{\mathbb R}$ the real parts of the momenta are no longer equal $\tilde p_1\neq\tilde p_2$. On this branch there are two possible solutions to the condition $x_1^-=x_2^+$  but only the one with $v_1>v_2$ corresponds to a bound state. This is the solution with $\tilde p_2>\tilde p_0>\tilde p_1$ which lies on the sheet ${\cal R}_{0,-1}$.\footnote{The corresponding branch with negative $\tilde p_{1,2}$ lies on the sheet ${\cal R}_{-1,-2}$.} 

We now consider the issue, relevant for the first branch, that there are two solutions with different values of $r$. For this branch (for both solutions) the total momentum $\tilde p_1 + \tilde p_2$ lies between $0$ and $2\tilde p_0$. For the second branch the total momentum is greater than $2 \tilde p_0$, and furthermore, there is only a single bound state solution for a given total momentum greater than $2\tilde p_0$. It therefore seems likely that there should also only be a single solution below this threshold so that there is only a single bound state for each value of the total momentum $\tilde p_1+\tilde p_2$. The likeliest explanation is that only the smaller root, denoted by the solid black line, corresponds to a bound state. The dashed line would then have to correspond instead to an anomalous threshold or the pole must be cancelled by a zero of the dressing phase in this region of parameter space. A definitive statement on this issue could only be made with a detailed investigation of the dressing phase and the kinematics required to produce an anomalous threshold in the form of a simple pole. An example of the latter, where a potential bound-state pole is actually an anomalous threshold, occurs in soliton scattering in the sine-Gordon theory \cite{Dorey:1996gd}.

\noindent{\bf Magnon kinematics}

\begin{figure}[ht]
\begin{center}
\begin{tikzpicture}[scale=1.2]
\draw[-] (0,0) -- (0,6) -- (6,6) -- (6,0) -- (0,0);
\draw[very thick] (0.6,0) .. controls (0.6,0.3) and (0.8,0.6) .. (0.8,0.9);
\draw[very thick] (0.8,0.9) .. controls (0.8,1.3) and (0.3,1.5) .. (0,1.5);
\filldraw[black] (0,1.5) circle (1pt);
\draw[very thick] (0,1.5) .. controls (0.3,1.8) and (0.6,2.5) .. (1.5,2.5);
\draw[very thick,densely dashed] (6,1.55) .. controls (4,1.7) and (3,2.5) .. (1.5,2.5);
\draw[very thick,densely dashed] (6,4.8) .. controls (4.5,5) and (3.5,5.4) .. (2.5,6);
\filldraw[black] (1.5,2.5) circle (1pt);
\draw[very thick,red] (1,0) .. controls (1,0.3) and (1.5,1) .. (1.5,2.5);
\draw[very thick,blue] (1,6) .. controls (1,5) and (1.5,4) .. (1.5,2.5);
\draw[dashed,black!50] (0,4.5) -- (6,4.5);
\draw[dashed,black!50] (0,1.5) -- (6,1.5);
\node at (3,-0.7) (a1) {$r$};
\node at (-1.2,3) (a2) {$\tilde p$};
\draw[-] (0,2.5) -- (0.1,2.5);
\node at (-0.4,2.5) (z1) {$\tilde p_0$};
\node at (-0.4,0) (c3) {0};
\node at (-0.4,6) (c4) {$\pi g$};
\node at (-0.5,1.5) (a7) {$\dfrac{\pi g}k$};
\node at (-0.8,4.5) (a8) {$\pi g -\dfrac{\pi g}k$};
\node at (3.1,3.1) (c1) {\bf magnon};
\node at (3.3,2.8) (l1) {$x_1^+=x_2^-$, $x_1^+\in{\mathbb R}$};
\node at (3.4,0.6) (c2) {{\bf soliton} $x_1^-=x_2^+$, $x_1^-\in{\mathbb R}$};
\node[blue] at (3.1,4.1) (c3) {\bf magnon};
\node[blue] at (3.3,3.8) (l2)  {$x_1^+=x_2^-$, $|x_1^+|=1$};
\node[blue] at (1.9,5.3) (a3) {$\tilde p_2$};
\node[red] at (2.2,2) (a4) {$\tilde p_1$};
\draw[->,blue] (a3) -- (1.2,5.3);
\draw[->,red] (a4) -- (1.6,2);
\draw[->] (c2) -- (0.8,0.6);
\draw[->,blue] (c3) -- (1.5,4.1);
\draw[->] (2.3,3.1) .. controls (2.3,3.1) and (1.6,3) .. (1.1,2.6);
\draw[->] (3.9,3.125) .. controls (5,3.125) and (5.2,4) .. (5.5,4.8);
\draw[->] (3.9,2.5) -- (4.2,2);
\end{tikzpicture}
\end{center}  
\caption{\small Schematic plot (values exaggerated for clarity) of the bound-state momenta of the 2 particles $p_1=\tilde p_1+ir$ and $p_2=\tilde p_1-ir$ for magnon kinematics. The solid black part of the curve corresponds to the bound state with $\tilde p_1=\tilde p_2\leq\tilde p_0$. This part of the curve is split into two branches corresponding to the soliton for $\tilde p_1=\tilde p_2<\frac{\pi g}k$ on which $x_1^-=x_2^+\in{\mathbb R}$ and the magnon for $\tilde p_1=\tilde p_2>\frac{\pi g}k$ on which $x_1^+=x_2^-\in{\mathbb R}$. The magnon branch then touches the third branch (also a magnon branch) at $x_1^+=1$ which continues with $x_1^+=x_2^-$ and $|x_1^+|=1$. The real parts of the momenta are no longer equal since $\tilde p_2>\tilde p_0$ (blue) and $\tilde p_1<\tilde p_0$ (red). On the fourth branch (also of magnon type), $\tilde p_1= \tilde p_2$, $\pi g-\frac{\pi g}{k}<\tilde p_1<\pi g$, and $-\pi g<\tilde E_1 =\tilde E_2 <0$.}
\label{f4}
\end{figure}
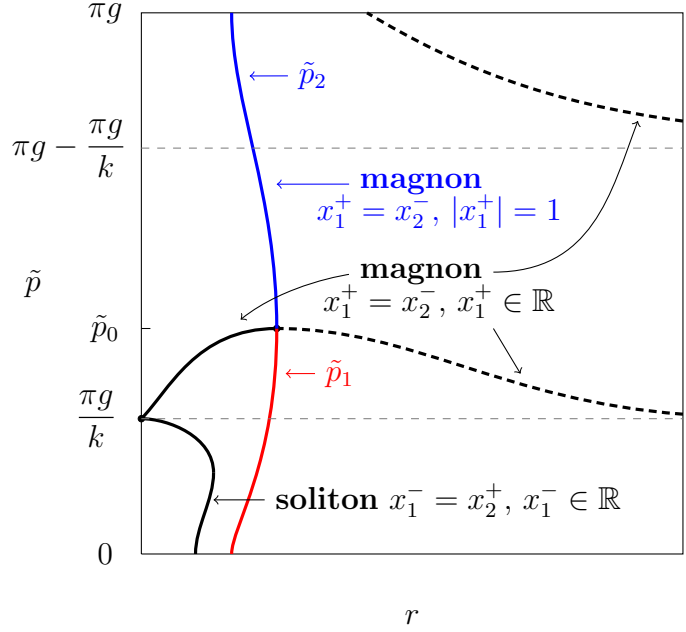

The case of magnon kinematics is more subtle as we illustrate in Fig.~\ref{f4}. In this case we have the parameterization
\EQ{
&U_1=e^{(i\tilde p_1-r)/2g}\ ,\qquad U_2=e^{(i\tilde p_2+r)/2g}\ ,\\
&V_1=e^{(i\tilde E_1-s)/2g}\ ,\qquad V_2=e^{(i\tilde E_2+s)/2g}\ .
}
There are four distinct branches of solutions. The branches denoted by solid black lines in Fig.~\ref{f4} lie on the sheet ${\cal R}_{0,0}$, while the dashed black line lies on the sheet ${\cal R}_{1,-1}$. The coloured lines lie on the intersection of the boundaries of these two sheets. The first branch has $\tilde p_1 = \tilde p_2<\pi g/k$ and we call it the ``soliton branch" since it corresponds to $x_1^-=x_2^+$ and a bound state in representation $\langle 0,1 \rangle^{\times 2}$. The other three branches have $\tilde p_2>\pi g/k$ and we call them  ``magnon branches" since they correspond to $x_1^+=x_2^-$ and a bound state in representation $\langle 1,0 \rangle^{\times 2}$. Recall that with magnon kinematics the momentum of states is restricted to the interval $|p|\leq g\pi$, and that the coproduct~\eqref{cop}  ensures the conservation of energy and momentum only modulo $2\pi g$. In fact, in Fig.~\ref{f4}, the states corresponding to the dashed line with $\pi g-\frac{\pi g}{k}<\tilde p_1= \tilde p_2<\pi g$ have energy and momentum\,\footnote{It is not difficult to show that in the limit $r\to\infty$ there are two solutions with
\EQ{
\tilde p_1= \tilde p_2\to \pm \frac{\pi g}{k}\,, \qquad \tilde E_1= \tilde E_2\to \frac{\pi g}{k}
}
and
\EQ{
\tilde p_1= \tilde p_2\to \pm (\pi g-\frac{\pi g}{k})\,, \qquad \tilde E_1= \tilde E_2\to  \frac{\pi g}{k}-\pi g
}
that correspond to the asymptotes of the dashed lines in Fig.~\ref{f4}. Notice that for the second solution $\tilde E_1= \tilde E_2$ turn out to be $<0$.
}
\EQ{
E=\tilde E_1+\tilde E_2 +2\pi g\,, \qquad p=\tilde p_1+\tilde p_2 \pm2\pi g\ .
}

As $\tilde p_{1}$ and $\tilde p_2$ increase from 0, the bound state corresponds to a soliton with $x_1^-=x_2^+$. On this branch $x_1^-\in{\mathbb R}$, while $\tilde p_1=\tilde p_2$ and $\tilde E_1=\tilde E_2$. However, this branch only exists for $\tilde p_1<\pi g/k$. At the end point $\tilde p_1=\pi g/k$, the imaginary part $r$ goes to zero. This point has an interesting interpretation. If we consider the $a^\text{th}$ bound state then it is easy to see that \eqref{dsr2} has the solution
\EQ{
U^2=V^2=q^a\quad\implies\quad E=p=a\frac{\pi g}k\ .
}
At this point the $a^\text{th}$ bound state is marginally unstable for decay into $a$ fundamental particles with $U=U_1U_2\cdots U_a$ and $V=V_1V_2\cdots V_a$ or equivalently
\EQ{
E_1=E_2=\cdots=E_a=\frac{\pi g}k\ ,\qquad p_1=p_2=\cdots=p_a=\frac{\pi g}k\ .
}
Hence, for 2-body scattering the point $\tilde p_{1,2}=\pi g/k$ is the point at which the bound state of 2 solitons in the representation $\langle0,1\rangle^{\times2}$ becomes un-bound and the bound state of 2 magnons in the representation $\langle1,0\rangle^{\times2}$ becomes bound as $\tilde p_{1,2}$ are increasing, or vice-versa when $\tilde p_{1,2}$ are decreasing. The picture we have established above means that the bound state with $a=2$ has two different branches as we anticipated earlier in section \ref{sec2} based on the dispersion relation. When the bound state momentum $p=\tilde p_1+\tilde p_2$ is less than $2\pi g/k$ the bound state is a soliton transforming in representation $\langle0,1\rangle^{\times2}$. On the contrary when the bound state momentum $p>2\pi g/k$ the bound state is a magnon transforming in representation $\langle1,0\rangle^{\times2}$.

As $\tilde p_1$ and $\tilde p_2$ increase beyond $\pi g/k$ we move onto one of the magnon branches with $x_1^+=x_2^-$. For $\tilde p_2<\tilde p_0$, we have $x_1^+\in{\mathbb R}$ which implies $\tilde p_1=\tilde p_2$ and $\tilde E_1=\tilde E_2$. The critical value $\tilde p_0$ corresponds to the point where the two magnon branches meet at $x_1^+=1$. As in the case of mirror kinematics there are two solutions for $r$ for a given $\tilde p_1$ shown by the solid and dashed lines and the same argument suggests that only the solid line corresponds to a bound state while the dashed line is an anomalous threshold. On the second magnon branch, $|x_1^+|=1$, there are two solutions and we pick the one with $v_1>v_2$ which implies $\tilde p_2>\tilde p_0>\tilde p_1$ with
\EQ{
e^{i(\tilde p_1-\tilde p_2)/g}=\frac{\frac1{x_1^{+*}}+\xi}{\frac1{x_1^+}+\xi}\ ,\qquad e^{i(\tilde E_1-\tilde E_2)/g}=\frac{1+x_1^+\xi}{1+x_1^{+*}\xi}\ .
}

The third magnon branch, with $\tilde p_1= \tilde p_2$ and $\pi g-\frac{\pi g}{k}<|\tilde p_1|<\pi g$, is obtained with $-\xi<x_1^+=x_2^-\in{\mathbb R}<0$. It corresponds to total momentum $0< |p|=2\pi g- 2|\tilde p_1|<\frac{2\pi g}{k}$, which overlaps with the range of momenta of the soliton branch. In this case, $-\pi g<\tilde E_1 =\tilde E_2 <0$ so that the total energy is $0<E=\tilde E_1+ \tilde E_2 + 2\pi g<\pi g$. Again, since we expect a single bound state for each value of the total momentum, we conjecture that this branch, which is the only one where the conservation of energy and momentum is non-standard, also gives rise to anomalous thresholds.

\section{Discussion}

One interesting conclusion of the analysis of \cite{Hoare:2011wr} is that the magnon-like S-matrix did not satisfy the correct crossing symmetry equation in the relativistic limit. However, the subtleties of the magnon case described above show that this conclusion was too hasty. In fact, in the relativistic limit, only the bound state below the threshold at $p=2\pi g/k$ is relevant and survives. Consequently the correct identification of the parameters $x^\pm=x(u\pm\frac i2)$ and the relativistic rapidity is $e^\theta=-q^{-iu}$ (where the sign implies that for real (physical) rapidities $q^{-iu}<0$, as required) and {\it not\/} $e^{-\theta}=q^{-iu}$ as claimed in \cite{Hoare:2011wr}. With this change the magnon S-matrix is consistent with crossing symmetry in the relativistic limit. So in this improved understanding in both the magnon and mirror theories the bound states that survive in the relativistic limit are the anti-symmetric ones $\langle 0,a-1\rangle^{\times2}$ which are identified with the solitons of the SSSSG theory. It is important to bear in mind that from the point-of-view of the SSSSG theory taking the solitons to transform in either  $\langle a-1,0\rangle^{\times2}$ or  $\langle 0,a-1\rangle^{\times2}$ is simply a matter of convention because the representations are isomorphic.

It is interesting that this assignment of representations implies that the identification of the $\msu(2)^{\oplus4}$ sub-algebras with the $\text{AdS}_5\times S^5$ geometry has to be re-arranged as one interpolates from the string limit to the relativistic limit. To appreciate this notice that in the string limit the states are associated to two of the four $\msu(2)$ symmetries. For instance the representation $ \langle a-1,0\rangle^{\times2}$ includes the states $(a,0,a,0)$, and the two $\msu(2)$'s associated to spin-$\frac a2$ factors are the ones associated to the $S^5$ part of the geometry. In the relativistic limit, the states in the $ \langle 0,a-1\rangle^{\times2}$ includes the states $(0,a,0,a)$. But from the point of view of the SSSSG theory, where the bound states are solitons, the two $\msu(2)$'s associated to spin-$\frac a2$ factors are also associated to the ``$S^5$'' part of the geometry.\footnote{In the SSSSG theory, the phrases ``$S^5$'' and ``$\text{AdS}_5$'' parts of the geometry refer to its origin as the Pohlmeyer reduction of superstring theory on $\text{AdS}_5 \times S^5$ \cite{Grigoriev:2007bu}. Solitons are naturally associated to the ``$S^5$'' part of the SSSSG theory where they are embeddings of the complex sine-Gordon soliton into the symmetric space sine-Gordon theory based on $S^5=\text{SO}(6)/\text{SO}(5)$~\cite{Hollowood:2011fq,Hollowood:2011fm}. If one tries to put a soliton in the ``$\text{AdS}_5$'' part it turns out to have singularities. In this sense it is a relative of a spiky string that goes to the boundary of $\text{AdS}_5$.} Therefore, if the interpolating theory can be given a geometrical setting then it must involve a kind of permutation $\text{AdS}_5\times S^5\leftrightarrow S^5\times \text{AdS}_5$. This is potentially an important clue for the search for a geometrical interpretation of the $q$-deformed theory.

\section*{Acknowledgements}

\noindent
BH is supported by ERC Advanced Grant No.\,290456 and would like to thank Arkady Tseytlin for enjoyable collaborations and useful discussions on related topics.

\noindent
TJH is supported in part by the STFC grant ST/G000506/1. 

\noindent
JLM is supported in part by MINECO (FPA2011-22594 and 
FPA2008-01177), Xunta de Galicia (Consejer\'\i a de Educaci\'on and INCITE09.296.035PR), the Spanish Consolider-Ingenio 2010
Programme CPAN (CSD2007-00042), and FEDER. He thanks Gleb Arutyunov for valuable discussions about his work.

\vspace{1cm}

\appendix
\appendixpage

\section{A similarity between (\texorpdfstring{$x^+,x^-,q,\xi$}{x+,x-,q,xi}) and (\texorpdfstring{$V^2,U^2,-\xi,-q$}{V2,U2,-xi,-q}).\label{appa}}
	
As discussed in section \ref{sec2} the fundamental particle states are labelled by two kinematical variables $x^\pm$ subject to the dispersion relation \eqref{p11}. In this appendix we find it convenient to rewrite this relation as 
\EQ{\label{dis1}
q^{-1}\left(x^+ + \frac1{x^+} + \xi + \frac1\xi\right) = q\left(x^- + \frac1{x^-} + \xi +\frac1\xi\right)\ .
}
The quantities $x^\pm$ are related to $U^2$ and $V^2$, which are simple functions of the energy and momentum, as in \eqref{jww} and the inverse relations are given in \eqref{xpm}. The dispersion relation can therefore be written in terms of $U$ and $V$ as in \eqref{dsr}. Writing this relation in terms of only the couplings $\xi$ and $q$ (using the definitions \eqref{couplings}) we find 
\EQ{\label{dis2}
\xi^{-1}\left(V^2 + \frac1{V^{2}} - q - \frac1q\right) = \xi \left(U^2 + \frac1{U^2} - q - \frac1q\right)\ .
}
This form is particularly appealing as we immediately see a similarity between the sets $(x^+,x^-,q,\xi)$ and $(V^2,U^2,-\xi,-q)$. Of course this is purely formal as $\xi \in [0,1]$, while $q = \exp(i\pi/k)$ is a phase. Furthermore, for the magnon and mirror theories the variables $x^\pm$ satisfy the reality condition
\EQ{
\left(x^+ + \frac1{x^+}\right)^* = x^- + \frac1{x^-} \ ,
}
while $U^2$ and $V^2$ satisfy
\EQ{
U^2+\frac1{U^2}\in \R \ , \qquad V^2 + \frac1{V^2} \in \R \ .
}

The relation \eqref{dis1} can be solved by introducing a new variable $u$ and defining the map $x(u)$ as in \eqref{sqr}. Again we rewrite this map in terms of the couplings $q$ and $\xi$
\EQ{\label{sqrb}
x + \frac1x + \xi +\frac1 \xi = (\xi^{-1} - \xi) \, q^{-2iu} \ .
}
The coefficient of $q^{-2iu}$ is greater than or equal to zero for $\xi \in [0,1]$. The dispersion relation \eqref{dis1} (or equivalently \eqref{p11}) is then solved by
\EQ{
x^\pm = x\Big(u \pm \frac i2\Big) \ .
}
Motivated by this we can similarly solve \eqref{dis2} by introducing a new variable $w$ and defining the map $W^2(w)$
\EQ{\label{comb1}
W^2 + \frac1{W^2} - q - \frac1q = \nu^2 \, \xi^{-2w}\ ,
}
where $\n$ is an arbitrary complex number.\footnote{As was done for the map $x(u)$ in \cite{Hoare:2011wr} this coefficient should be fixed by considering the string and relativistic limits.} The dispersion relation \eqref{dis2} is then solved by
\EQ{\label{comb2}
V^2 = W^2(w - \frac12) \ , \qquad U^2 = W^2(w + \frac12) \ .
}

A map between $\xi^{-2w}$ and $q^{-2iu}$ can be found by considering the relation
\EQ{\label{relll}
(\xi^{-1} - \xi)\, q^{-2iu} = q^{-1}\left(x^+ + \frac1{x^+} + \xi + \frac1\xi\right)\ ,
}
and substituting in for $x^+$ and $1/x^+$ using the two different expressions in \eqref{xpm}. The resulting map is then given by
\EQ{\label{uwmap}
q^{-2iu} = \frac{(UV - (UV)^{-1})^2}{(\xi^{-1}-\xi)\, \nu^2 \, \xi^{-2w}} \ .
}

This similarity between $(x^+,x^-,q,\xi)$ and $(V^2,U^2,-\xi,-q)$, and the map between $u$ and $w$ \eqref{uwmap} is particularly appealing and may be indicating some additional underlying structure of this rather special interpolating theory. Here we will use it to prove that for magnon kinematics, $U^2 = \exp(i p/g)$ and $V^2 = \exp(i E/g)$, $q^{-2iu}$ is negative for $|p|<\pi g/k$ and positive for $|p|>\pi g/k$ as claimed in section \ref{sec2}---see equation \eqref{claim}.

Combining \eqref{comb1}, \eqref{comb2} and \eqref{uwmap} it is relatively simple to see
\EQ{\label{rrpre}
\xi\left(U^2 + U^{-2} - q - q^{-1}\right) = \frac{((UV) - (UV)^{-1})^2}{(\xi^{-1} - \xi) \, q^{-2iu}}\ .
}
Substituting in for $V$ and $U$ in terms of energy and momentum (and also for $q$ in terms of $k$) implies the following relation\,\footnote{Note that the map $W(w)$ does not actually need to be introduced to find this relation---it can be found by directly substituting in for $x^+$ in \eqref{relll}.}
\EQ{\label{rr}
q^{-2iu} = \frac{2\sin^2(\frac{E+p}{2g})}{(1-\xi^2)(\cos \frac{\pi}{k}-\cos \frac{p}{g})}
         = \frac{\sin(\frac{p+E}{2g})}{\sin(\frac{p-E}{2g})}\ .
}
Recalling that $1 - \xi^2 > 0$ we immediately see that for $|p|<E$, that is $|p|<\pi g/k$, $q^{-2iu}<0$ and for $|p|>E$, that is $|p|>\pi g/k$, $q^{-2iu}>0$ as claimed. Notice that in the relativistic limit $g\to\infty$ the variable $u$ is related to the usual relativistic rapidity via $q^{-2iu}=-e^{2\theta}$.

This relation can be easily generalized to the representations corresponding to the bound states; namely, $\langle a-1,0\rangle^{\times2}$ or $\langle 0,a-1\rangle^{\times2}$. The particle states in these representations are also labelled by two kinematical variables $x^\pm$ that, in this case, are subject to the dispersion relation~\eqref{bsdr}, which can also be solved in terms of the map $x(u)$ defined in~\eqref{sqr} (or~\eqref{sqrb}) as follows
\EQ{\label{bspp}
x^\pm = x\Big(u\pm \frac{ia}{2}\Big)\ .}
Then, using the relations~\eqref{bsuv}, it is not difficult to  check that the generalization of the relation~\eqref{rr} to the case of the bound state representations with magnon kinematics is simply 
\EQ{\label{rrbs}
q^{-2iu} = \frac{ 2\sin^2(\frac{E+p}{2g})}{(1-\xi^2)(\cos \frac{\pi a}{k}-\cos \frac{p}{g})}
         = \frac{\sin(\frac{p+E}{2g})}{\sin(\frac{p-E}{2g})}\ .
}
Note that the final expression in the above does not depend on $a$ and is therefore universal. This confirms that there are two branches, a soliton branch with $|p|<E$, that is $|p|<a\pi g/k$, for which $q^{-2iu}<0$, and a magnon branch with $|p|>E$, that is $|p|>a\pi g/k$, for which $q^{-2iu}>0$. 

Since the physical values of the momentum correspond to $|p|\leq g\pi$, it is an interesting observation that (for $k \in \mathbb{Z}$) when $a = k$ only the soliton branch remains (recall that in the string limit $k \to \infty$, hence this effect disappears). However, in this case $q^a=1$, i.e. $q$ is a root of unity, and the structure of the corresponding representations requires further understanding.

\end{document}